\documentclass[11pt]{article}

\usepackage{times}

\usepackage{graphicx}

\begin{document}
\vspace*{-40pt}
\begin{flushright}
\textsf{A\&G {\bf 47} (2006) 4.30--4.33}
\end{flushright}

\vspace*{20pt}
\begin{center}
{\Large COSMOLOGICAL MODEL SELECTION}

\vspace*{12pt}
{\large {\em Andrew R Liddle, Pia Mukherjee and David Parkinson}}\\
Astronomy Centre, University of Sussex, Brighton BN1 9QH
\end{center}

\begin{abstract}
Model selection aims to determine which theoretical models are most
plausible given some data, without necessarily asking about the
preferred values of the model parameters. A common model selection
question is to ask when new data require introduction of an additional
parameter, describing a newly-discovered physical effect. We review
several model selection statistics, and then focus on use of the
Bayesian evidence, which implements the usual Bayesian analysis
framework at the level of models rather than parameters. We describe
our \emph{CosmoNest} code, which is the first
computationally-efficient implementation of Bayesian model selection
in a cosmological context. We apply it to recent WMAP satellite data,
examining the need for a perturbation spectral index differing from
the scale-invariant (Harrison--Zel'dovich) case.
\end{abstract}

\section{Introduction}

Cosmologists are becoming very good at determining the parameters of
the Universe. Within the last few years observational results,
exemplified by the microwave background anisotropy measurements by the
Wilkinson Microwave Anisotropy Probe (WMAP), have introduced precision
into cosmological modelling. Considerable sophistication is now
required both in deriving theoretical predictions from models and in
carrying out data analysis procedures able to squeeze the best from
the data.

Within a cosmological model, the parameters indicate the importance of
different effects. For instance, they describe the relative amounts of
different types of material in the Universe, the geometry and
expansion rate of the Universe, and the properties of the initial
irregularities in the Universe which led to the formation of
structure. Such parameters are not predicted by fundamental theories,
but rather must be fit from data in order to decide which combination,
if any, is capable of describing our Universe. A variety of
cosmological data are currently well fit by a model of the Universe
that is homogeneous, isotropic and spatially flat, contains cold dark
matter in greater proportions than ordinary baryonic matter, and in
which tiny initial perturbations evolved under Einstein's theory of
General Relativity into the structures of today. Current data
constrain the parameters of this model rather well, many at the 10\%
level or better.

However, the presence of good data leads to a different problem, one
of knowing when to stop fitting. Two different and competing models of
the Universe may explain the data equally well, so how do we choose
between them? The solution is one proposed by William of Occam, that
the simpler model should be preferred. This is known as \emph{Occam's
razor}. So a complicated model that explains the data slightly better
than a simple one should be penalized for the extra parameters it
introduces, because the extra parameters bring with them a lack of
predictability. On the other hand, if a model is too simple, and
cannot fit certain data well, then it can be discarded.  This is a
rather common type of statistical problem, both in cosmology and in
other fields of astrophysics: each available parameter within a model
describes some piece of physics that might be relevant to our
Universe, but until measurements are made we don't know which.

Cosmological \emph{model selection} refers to comparing different
model descriptions of the data. It doesn't care particularly about the
actual values of parameters, but rather aims to determine which {\em
set} of parameters gives the preferred fit to observational data.

\section{Why model select?}

Model selection is an extremely widespread challenge throughout
science; how do you fit to data when you are unsure about the set of
parameters that you should be deploying. You cannot just include every
parameter you can think of in a fit to data, because inclusion of
extra unnecessary parameters worsens the determination of those that
are essential, so that very soon you can end up learning nothing about
anything. Moreover, you can't just use goodness of fit to the data,
because typically inclusion of a new parameter will improve the
goodness of fit even if that parameter has absolutely no actual
relevance to the Universe. Typical attempts to avoid these problems
involve {\it ad hoc} criteria such as `chi-squared per degrees of
freedom' arguments or the `likelihood ratio test', in which arbitrary
thresholds have to be invoked to decide which way the verdict is
supposed to go. Model selection aims to put this practice on a firmer
footing.

Model selection problems are ones in which the parameter set necessary
to describe a given dataset is unknown, the question typically being
whether new data justifies inclusion of a new physical parameter. Many
of the most pressing questions in astrophysics are of this form.
Cosmological examples would include whether the spatial curvature is
non-zero, whether the dark energy density evolves, and whether the
initial perturbation spectrum has an amplitude which varies with
length scale.

\section{Model selection statistics}

The generic purpose of a model selection statistic is to set up a
tension between the predictiveness of a model (for instance indicated
by the number of free parameters) and its ability to fit observational
data. Oversimplistic models offering a poor fit should of course be
thrown out, but so should more complex models which offer poor
predictive power.

There are two main types of model selection statistic that have been
used in the literature so far. {\bf Information criteria} look at the
best-fitting parameter values and attach a penalty for the number of
parameters; they are essentially a technical formulation of
`chi-squared per degrees of freedom' arguments. By contrast, the {\bf
Bayesian evidence} applies the same type of likelihood analysis
familiar from parameter estimation, but at the level of models rather
than parameters. It depends on goodness of fit across the entire model
parameter space.

Here we discuss three possible statistics. In each case, the statistic
is a single number that is a property of the model, and having
computed it the models can be placed in a rank-ordered list.
\begin{description}
\item[Akaike Information Criterion (AIC):] This was derived by
Hirotugu Akaike in 1974, and takes the form
\begin{equation}
{\rm AIC} = -2\ln {\cal L}_{\rm max} + 2k \,,
\end{equation}
where ${\cal L}$ is the likelihood ($-2\ln {\cal L}$ is often called
$\chi^2$ though it generalizes it to non-gaussian distributions) and
$k$ is the number of parameters in the model. The subscript `max'
indicates that one should find the parameter values yielding the
highest possible likelihood within the model.  It is obvious that this
second term acts as a kind of `Occam factor'; initially as parameters
are added the fit to data improves rapidly until a reasonable fit is
achieved, but further parameters then add little and the penalty term
$2k$ takes over. The generic shape of the AIC as a function of number
of parameters is therefore a rapid fall, a minimum, and then a
rise. The preferred model sits at the minimum.

The AIC was derived from information-theoretic considerations,
specifically an approximate minimization of the Kullback--Leibler
information entropy which measures the distance between two
probability distributions.

\item[Bayesian Information Criterion (BIC):] This was derived by
Gideon Schwarz in 1978, and strongly resembles the AIC. It is given by
\begin{equation}
{\rm BIC} = -2\ln {\cal L}_{\rm max} + k \ln N \,,
\end{equation}
where $N$ is the number of datapoints. Since a typical dataset
will have $\ln N >2$, the BIC imposes a stricter penalty against extra
parameters than the AIC.

It was derived as an approximation to the Bayesian evidence, to be
discussed next, but the assumptions required are very restrictive and
unlikely to hold in practice, rendering the approximation quite crude.

\item[Bayesian evidence:] The Bayesian evidence looks rather
different, being defined as
\begin{equation}
E = \int {\cal L}(\theta) {\rm Pr}(\theta) d\theta \,.
\end{equation}
Here $\theta$ is the vector of parameters of the model, and ${\rm
Pr}(\theta)$ is the prior distribution of those parameters {\em
before} the data were obtained. The prior is an essential part of the
definition of a model, upon which the evidence will ultimately depend,
and might for instance be a set of ranges within which parameters are
assumed to be uniformly distributed.

The evidence of a model is thus the average likelihood of the model in
the prior. Unlike the statistics above, it does not focus on the
best-fitting parameters of the model, but rather asks ``of all the
parameter values you thought were viable before the data came along,
how well on average did they fit the data?''. Literally, it is the
likelihood of the model given the data. Given Bayes' theorem
\begin{equation}
P(M|D)=\frac{P(D|M)P(M)}{P(D)} \,.
\end{equation}
(here M is the model, D is the data, and the vertical bar is read as
`given'), the evidence \mbox{$E \equiv P(D|M)$} updates the prior model
probability $P(M)$ to the posterior model probability $P(M|D)$,
i.e.~the probability of the model given the data.

The evidence rewards predictability of models, provided they give a
good fit to the data, and hence gives an axiomatic realization of
Occam's razor. A model with little parameter freedom is likely
to fit data over much of its parameter space, whereas a model which
could match pretty much any data that might have cropped up will give
a better fit to the actual data but only in a small region of its
larger parameter space, pulling the average likelihood down.

The evidence is also known as the marginalized likelihood or, more
accurately, the model likelihood. The ratio of evidences for two
models is known as the Bayes factor.

\end{description}

Of these statistics, we would advocate using, wherever possible, the
Bayesian evidence which is a full implementation of Bayesian inference
and can be directly interpreted in terms of model probabilities. It
is computationally challenging to compute, being a highly-peaked
multi-dimensional integral, but recent algorithm development has made
it feasible in cosmological contexts. We discuss it further in the
next section.

If the Bayesian evidence cannot be computed, the BIC can be deployed
as a substitute. It is much simpler to compute as one need only find
the point of maximum likelihood for each model. However interpreting
it can be difficult. Its main usefulness is as an approximation to the
evidence, but this holds only for gaussian likelihoods and provided
the datapoints are independent and identically distributed. The latter
condition holds poorly for the current global cosmological dataset,
though it can potentially be improved by binning of the data hence
decreasing the $N$ in the penalty term.

The AIC has been widely used outside astrophysics, but is of debatable
utility. Sometime after it was first derived, it was shown to be
`dimensionally inconsistent', a statistical term meaning that it is
not guaranteed to give the right result even in the limit of infinite
unbiased data. It may however be useful for checking the robustness of
conclusions drawn using the BIC. The evidence and BIC are
dimensionally consistent.

\section{Computing and interpreting the evidence}

Computing the evidence in realistic problems is challenging,
particularly in cosmology where evaluating theoretical predictions at
just a single parameter point requires several seconds of CPU time
with state-of-the-art codes such as \texttt{cmbfast} or \texttt{camb}.
Markov chain Monte Carlo (MCMC) methods are now commonplace in
cosmological parameter estimation, and efficiently trace the posterior
probability distribution of the parameters of a model in the vicinity
of the best-fit region.  However a different sampling strategy is
needed to evaluate the evidence. It can receive a large contribution
from the tails of the posterior distribution of the parameters,
because even though the likelihoods there are small, this region
occupies a large volume of the prior probability space.  Therefore the
sampling strategy must effectively sample the entire prior volume to
evaluate the integral (Eqn 3) accurately.  Until recently, the best
available strategy for evidence calculation, known as thermodynamic
integration or simulated annealing, required around $10^7$ likelihood
evaluations for an accurate answer for a five-parameter cosmological
model, placing the problem at the limit of current supercomputer
power.

Fortunately, a powerful new algorithm for evidence evaluation, known
as {\bf nested sampling}, was recently invented by John Skilling
(2004). At Sussex we have implemented this algorithm for cosmology in
a code named \texttt{CosmoNest}, which we recently made publically
available. It has proven to be one to two orders of magnitude more
efficient than thermodynamic integration, meaning that evidence
calculations can now be run on a small computing cluster.

\begin{figure}[t]
\begin{center}
\includegraphics[width=0.5 \linewidth]{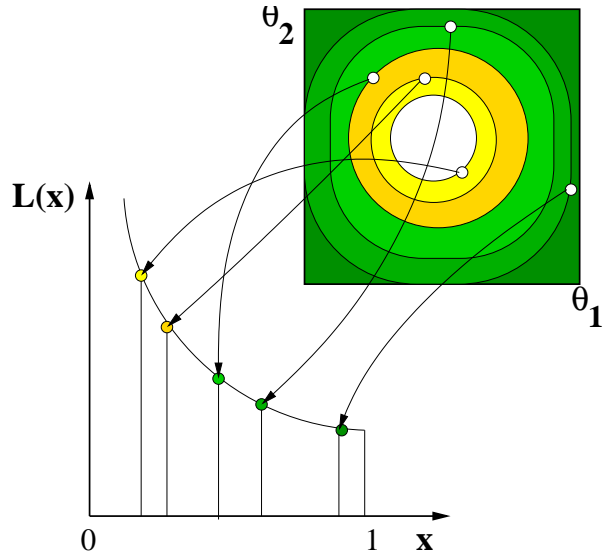}
\caption{A schematic of the nested sampling algorithm. The
two-dimensional parameter space is shown at the top right. The points
within it are considered to represent contours of constant likelihood,
which sit within each other like layers of an onion (there is however
no need for them to be simply connected). The volume corresponding to
each thin shell of likelihood is computed by the algorithm, allowing
the integral for the evidence to be accumulated as shown in the
graph.}
\label{fig:nested}
\end{center}
\end{figure}

To set up the algorithm, the evidence integral is first recast as a
one-dimensional integral in terms of the prior mass $X$, where $dX =
\mathrm{Pr}(\theta)\,d\theta$ with $X$ running from 0 to 1. [A mental
image to accompany this is to consider the prior parameter space as a
cube, and to smash it with a large hammer. The fragments are then
arranged in a line in order of increasing likelihood.] The algorithm
samples the prior a large number of times, assigning a `prior mass' to
each sample.  The samples are ordered by likelihood, and the
integration follows as the sum of the sequence,
\begin{equation}
E = \int L(X)dX = \sum_{j=1}^m E_j\,, \quad
E_j=\frac{L_j}{2}(X_{j-1}-X_{j+1}) \,.
\end{equation}
This is shown in Figure~\ref{fig:nested}.

\begin{figure}[t]
\begin{center}
\includegraphics[width=0.8 \linewidth]{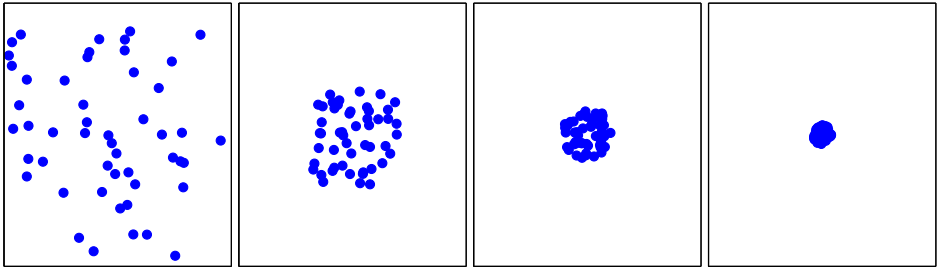}
\caption{A sequence of snapshots of a toy-model evidence calculation,
using a two-dimensional gaussian likelihood. As the computation
progresses the cluster of points, initially distributed throughout the
parameter space, drifts towards the region of high likelihood at the
centre.}
\label{fig:timeseries}
\end{center}
\end{figure}

In order to compute the integral accurately the prior mass is
logarithmically sampled. We start by randomly placing a set of $N$
points within the prior parameter space, where in a typical
cosmological application $N \simeq 300$. We then iteratively discard
the lowest likelihood point $L_j$, replacing it with a new point
uniformly sampled from the remaining prior mass (i.e.~with likelihood
greater than $L_j$).  Each time a point is discarded the prior mass
remaining, $X_j$, shrinks by a factor that is known probabilistically,
and the evidence is incremented accordingly. In this way the algorithm
works its way towards the higher likelihood regions. The process is
illustrated in Figure \ref{fig:timeseries}. Additional details of the
algorithm are in Mukherjee, Parkinson \& Liddle (2006a). The algorithm
is simple, works accurately even in high dimensions, and should be
generally applicable in a number of areas even outside of
astrophysics.

Although the evidence gives a rank-ordered list of models, it is still
necessary to decide how big a difference in evidence is needed to be
significant. If the prior probabilities of the models are assumed
equal, the difference in log(evidence) can be directly interpreted as
the relative probabilities of the models after the data.  Even if
people disagree on the relative prior probabilities, they will all
agree on the direction in which the data, represented by the evidence,
has shifted the balance. The usual interpretational scale employed is
due to Sir Harold Jeffreys (from his classic 1961 book `Theory of
Probability'), which, given a difference $\Delta \ln E$ between the
evidences $E$ of two models, states that
\begin{center}
\begin{tabular}{|c|l|}
\hline
$\Delta \ln E < 1$ & Not worth more than a bare mention.\\
$1 < \Delta \ln E < 2.5$ & Significant.\\
$2.5 < \Delta \ln E < 5$ & Strong to very strong.\\
$5 < \Delta \ln E$ & Decisive.\\
\hline
\end{tabular}
\end{center}
In practice we find the divisions at 2.5 (corresponding to posterior
odds of about 13:1) and 5 (corresponding to posterior odds of about
150:1) the most useful. 

When should model selection be deployed? If the data indicates
something strongly enough, it doesn't really matter how the
statistical analysis is done. The main zone of interest is where a new
parameter is `detected' at between two and four `sigmas' via parameter
estimation techniques. These overestimate the significance of a
detection because they ignore model dimensionality, and there is a
well-known (in the statistics literature anyway) phenomenon called
Lindley's paradox, whereby model selection considerations can overturn
an analysis based on a `number of sigmas' argument. A nice discussion
of Lindley's paradox is given in Trotta (2005).

One of the bugbears of Bayesian methods is the requirement to specify
priors explicitly, with the evidence depending on the choice of
priors. If the data have low informative content (technically defined
via the ratio of prior and posterior parameter volumes), this can be a
serious issue, but it becomes less so if the data are constraining so
that the posterior is well localized within any conceivable prior. In
that case the evidence becomes proportional to the prior volume, and
quite a substantial change in volume is needed to move models
significantly around the Jeffreys scale.

\section{Applications of model selection}

There are several areas of application of model selection techniques,
the main two being as follows:
\begin{description}
\item[Application to data:] With real data, one can assess the
viability of different models under consideration. In this case one
simply computes the evidence for each model of interest and ranks them.
\item[Model selection forecasting:] This application aims to compare
the power of different experiments before they are carried out.  Many
proposed experiments seek to answer model selection questions, but
their capabilities are often quantified using parameter estimation
projections, such as Fisher matrix forecasting. For instance, a dark
energy experiment may be advertized as able to measure the equation of
state parameter $w$ with an uncertainty of $\pm 0.05$, the aim being
to detect deviations of $w$ from $-1$, which characterizes the
cosmological constant or vacuum zero-point energy. One can instead
forecast experiments' ability to carry out model selection tests. In
this case data must be simulated for a range of different assumed
models, in order to investigate where in the available parameter space
a given experiment can make a strong or decisive model comparison
between a dynamical dark energy model and the cosmological
constant. This gives a powerful tool for comparing the statistical
power of competing experiments. It should also be possible to extend
this concept to survey optimization, whereby one tunes survey
parameters to optimize the ability to carry out model selection tests,
but it is less clear that this will be fruitful.
\end{description}
We have extensively discussed the philosophy of model selection
forecasting, with specific application to dark energy experiments, in
Mukherjee et al.~(2006b). In Pahud et al.~(2006) we applied these
ideas to determination of the nature of the primordial power spectrum
of density perturbations, focussing on the ability of the Planck
Satellite mission to perform model selection of this type. In this
article we will focus on applications to real data.

\subsection{A toy model}

To help understand what is going on, we can carry out a simple toy
model investigation into the spatial curvature of the
Universe. According to the three-year data from WMAP (henceforth
WMAP3), the total density, in units of the critical density, is
$\Omega = 1.003 \pm 0.015$ (where we took the liberty of symmetrizing
the uncertainty and where the Hubble Key project determination of the
Hubble parameter is also used). Given this, how likely is it that the
Universe is flat? For simplicity we'll assume a gaussian likelihood
corresponding to this measurement, and ignore dependence on other
parameters, so we have
\begin{equation}
{\cal L} = {\cal L}_0 \exp \left[ - \frac{\left(\Omega -
  1.003\right)^2}{2 \times 0.015^2} \right]
\end{equation}
We also have to choose a prior range for $\Omega$; let's say
$0.1<\Omega < 2$ representing some plausible range people might have
considered long before precision data emerged. Now the calculation,
remembering that the evidence is just the average likelihood over the
prior.
\begin{description}
\item[Flat model:] We just have to evaluate the likelihood at $\Omega
  = 1$. It is $E({\rm flat}) = 0.98 {\cal L}_0$.
\item[Curved model:] Now we have to integrate the likelihood over the
  prior, being sure to normalize the prior properly. This gives
  $E({\rm curved}) =  0.02 {\cal L}_0$.
\end{description}
The conclusion is that, under these assumptions, the flat model is
preferred at odds of approximately 50:1.

That example was pretty boring, since $\Omega=1$ lies almost in the
middle of the measured range. But suppose the result had been $\Omega
= 1.045 \pm 0.015$, a putative three-sigma detection of spatial
curvature. The evidence for the curved model is unchanged (it doesn't
care what the measured value is provided it is well within the prior)
while that of the flat model shrinks. Nevertheless, the end result is
an odds ratio of only 2:1 in favour of the curved model. In this case,
three-sigma is nowhere near enough to convincingly indicate that space
is curved.  Physically, the evidence is allowing for it being {\it a
priori} very unlikely that $\Omega$ could be so close to one as to
give such a low-confidence `detection', yet still not be equal to one.
Put another way, the $\Delta \ln E$ would be 0.7 which according to
Jeffreys is hardly worth mentioning.

\begin{figure}[t]
\begin{center}
\includegraphics[width=0.8 \linewidth]{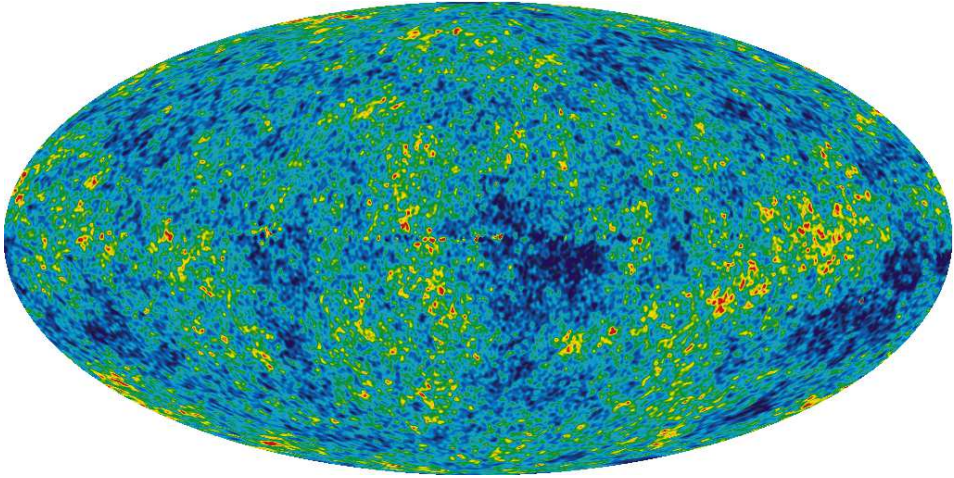}
\caption{A composite map of the cosmic microwave background (CMB) as
measured by three years of WMAP observations. The colour scale
indicates the CMB temperature and the whole sky is shown in
Hammer--Aitoff projection in galactic coordinates.  Measurements at
different frequencies have used to model out non-CMB contributions,
particularly galactic emission. [Image courtesy NASA/WMAP Science
Team.]}
\label{fig:ilc}
\end{center}
\end{figure}

\subsection{Real cosmology}

\begin{figure}[t!]
\begin{center}
\includegraphics[width=0.7 \linewidth]{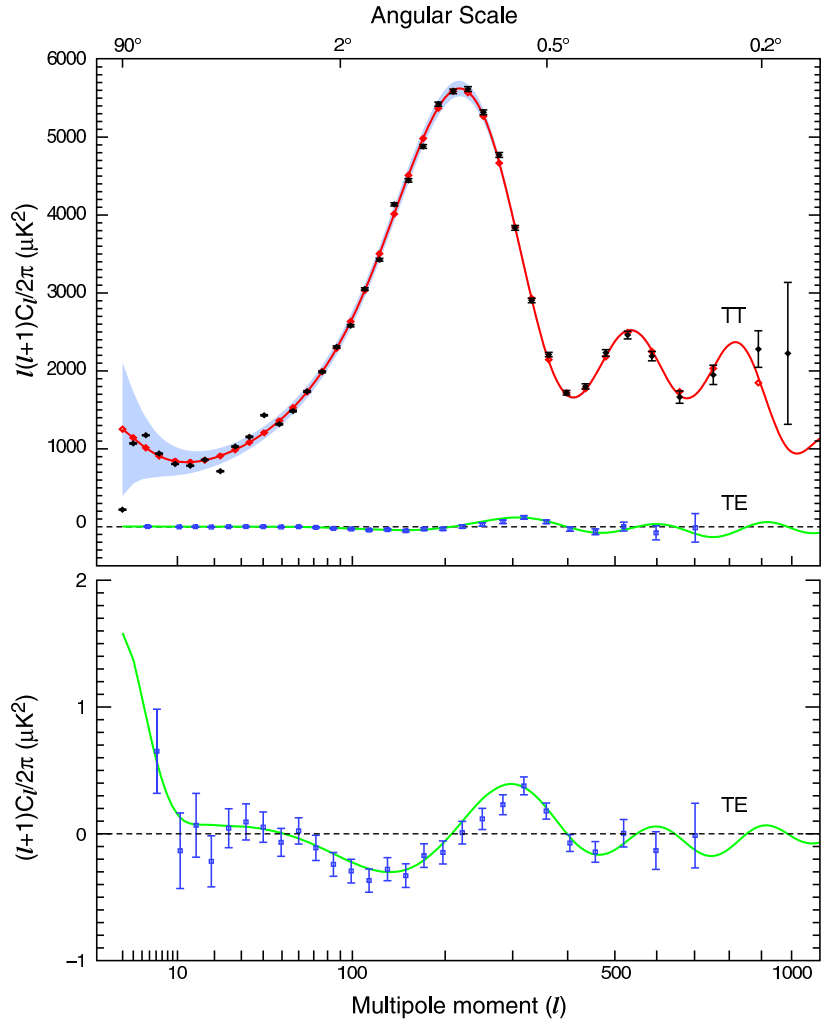}
\caption{Upper panel: the CMB temperature power spectrum measurements
from WMAP3 (black dots, with error bars) shown against a model power
spectrum that provides a good fit (red curve).  Bottom panel: the CMB
temperature--polarization cross-correlation measurements from WMAP
(blue dots with error bars) shown against the predictions of the same
model as above (green curve). [Image courtesy NASA/WMAP Science
Team.]}
\label{fig:spectrum}
\end{center}
\end{figure}

Now we turn to a real cosmological example. The new three-year data
from WMAP (see Figures~\ref{fig:ilc} and \ref{fig:spectrum}) is for
the most part uncontroversial from a model selection point of view,
with parameters either being definitely required or clearly
unnecessary. The exception is the scale dependence of primordial
density perturbations, defined by the spectral index $n_{{\rm S}}$.
These perturbations are usually considered to have been generated by
inflation, a period of rapid acceleration in the early Universe. As
well as solving some of the problems with the traditional hot big bang
model, inflation also generically predicts the kind of observations
that we now see. The many models of inflation predict a wide range of
possible values for $n_{{\rm S}}$, which one should then try and fit
from the data.

However, a decade before inflation was invented, Harrison and
Zel'dovich independently proposed that $n_{{\rm S}}$ should be
precisely one, corresponding to perturbations whose amplitude is
independent of scale. At least until this year's publication of the
three-year WMAP data, the Harrison--Zel'dovich spectrum always gave a
good fit to existing data. From a model selection perspective it
benefits from having one parameter less than a model where $n_{{\rm
S}}$ varies, and indeed we showed in a paper last year predating WMAP3
that the Harrison--Zel'dovich model had the highest evidence, though
other models including varying $n_{{\rm S}}$ were not strongly
excluded.

WMAP3 gave, for the first time, indications that $n_{{\rm S}}$ might
be less than one, with their main paper quoting the results (from
WMAP3 data alone) $n_{{\rm S}}= 0.951^{+0.015}_{-0.019}$, which thus
appears to be over 3-sigma away from unity. A similar result is found
when WMAP data is combined with other independent datasets, such as
the power spectrum of the large-scale distribution of galaxies and the
redshift--luminosity relation of distant type Ia supernovae.  So far,
parameter estimation analyses performed on available data taken
together seem to indicate that $n_{{\rm S}} \neq 1$ at about 3 to
4-sigma.

This significance level is exactly where Lindley's paradox is at its
strongest, making the use of model selection techniques imperative, as
acknowledged in the WMAP3 papers. We have carried out such an
analysis. We chose a prior on $n_{{\rm S}}$ uniform between 0.8
between 1.2; most inflationary models give $n_{{\rm S}}$ in this range
and this is what was believed to be the possible range for it before
the data came along.  Evidences were computed using
\texttt{CosmoNest}, with the calculations taking a few days on a
multi-processor cluster.

According to our model selection analysis, the evidence for the
$n_{{\rm S}}$ varying model is significant, but not strong or
decisive.  WMAP3 data on its own gives a Bayes factor of only $0.34\pm
0.26$, indicating that this data alone is unable to distinguish the
two models.  When WMAP3 data are used together with external data sets
we estimate a $\Delta \ln E$ of $1.99\pm 0.26$, corresponding to an
odds ratio of 8 to 1 in favour of the $n_{{\rm S}}$ varying model.
Adding the external datasets improves the constraining power on
$n_{{\rm S}}$, as they significantly extend the scales over which the
primordial power spectrum affects the data. Nevertheless, the support
for varying $n_{{\rm S}}$ is clearly tentative rather than compelling.

There is additional reason for some caution at present because there
may be residual systematics in the data that could affect our
conclusions regarding $n_{{\rm S}}$; the evidence calculation concerns
statistical uncertainties only. For example, the effect of varying
$n_{{\rm S}}$ in determining the power spectra shown in
Figure~\ref{fig:spectrum} is somewhat degenerate with the signature of
the relatively recent reionization of the Universe, which is mainly
inferred from polarization data which is difficult to handle. There
are also uncertainties associated with the modelling of the instrument
beam profiles, and in whether one should attempt to model out a
possible contribution to the CMB anisotropies from the
Sunyaev--Zel'dovich effect. The situation will be improved with higher
signal-to-noise data from additional years of WMAP observations and
future experiments.

\section{Conclusion}

Many of the most interesting cosmological questions are ones of model
selection, not parameter estimation.  With the growing precision of
cosmological data, it is imperative to deploy proper model selection
techniques to extract the best robust conclusions from data.

Application to the post-WMAP3 cosmological data compilation continues
to indicate that the data can be well fit by quite minimal
cosmological models. Five fundamental parameters are definitely
required, and WMAP3 has provided suggestive indications that a sixth,
the density perturbation spectral index, needs to be added to the
set. According to the Bayesian evidence, however, the case for
inclusion of the spectral index has yet to become compelling.

As the data improve in sensitivity we expect new model selection based
questions to be both raised and answered in the next decade.  These
may be about the nature of dark energy, the model for reionization,
the nature of inflation, the case for primordial gravitational waves,
or the nature of cosmic topology.  Model selection, of course, will
have further applications in astrophysics and beyond.

\section*{Acknowledgments}

The authors were supported by PPARC. We thank Pier Stefano Corasaniti,
Mike Hobson, Andrew Jaffe, Martin Kunz, C\'edric Pahud, John Peacock,
John Skilling, and Roberto Trotta for discussion relating to these
ideas.

\texttt{CosmoNest} is available for download at {\tt
http://www.cosmonest.org}.

\section*{Bibliography}

{\bf Jeffreys H} 1961, {\it Theory of Probability}, 3rd edition [OUP,
Oxford]. 

\noindent {\bf Liddle A R} 2004, {\it MNRAS} {\bf 351} L49-L53

\noindent {\bf Mukherjee P, Parkinson D, Liddle A R} 2006a, {\it ApJ}
{\bf 638} L51-L54 

\noindent {\bf Mukherjee P, Parkinson D, Corasaniti P S, Liddle A R,
Kunz M} 2006b, {\it MNRAS} {\bf 369}, 1725--1734

\noindent {\bf Parkinson D, Mukherjee P, Liddle A R} 2006, Phys. Rev. D
	{\bf 73}, 123523

\noindent {\bf Pahud C, Liddle A R, Mukherjee P, Parkinson D} 2006,
	Phys. Rev. D {\bf 73}, 123524

\noindent {\bf Skilling J} 2004, in {\it Bayesian Inference and Maximum
Entropy Methods in Science and Engineering} ed. R. Fischer et al.\
[Amer.\ Inst.\ Phys., conf.\ proc.,] {\bf 735} 395

\noindent {\bf Trotta R} 2005, {\it astro-ph/0504022}

\end{document}